\documentclass[10pt,a4paper]{article}

\usepackage[a4paper,top=17mm,bottom=18mm,left=16mm,right=16mm,headsep=5mm,footskip=9mm]{geometry}
\usepackage[T1]{fontenc}
\usepackage{libertinus}
\usepackage{textcomp}

\usepackage{microtype}
\usepackage{graphicx}
\graphicspath{{figures/}{manuscript/latex/figures/}}
\usepackage{booktabs}
\usepackage{array}
\usepackage{tabularx}
\usepackage{adjustbox}
\usepackage{float}
\newcolumntype{Y}{>{\centering\arraybackslash}X}
\usepackage[table]{xcolor}
\usepackage{caption}
\usepackage{titlesec}
\usepackage{enumitem}
\usepackage{fancyhdr}
\usepackage{dblfloatfix}
\usepackage{placeins}
\usepackage{balance}
\usepackage{xurl}
\usepackage[hidelinks]{hyperref}

\definecolor{PrismNavy}{HTML}{17324D}
\definecolor{PrismTeal}{HTML}{007C83}
\definecolor{PrismGold}{HTML}{C7922E}
\definecolor{PrismPale}{HTML}{EEF7F7}
\definecolor{PrismRule}{HTML}{C9D6DA}
\hypersetup{colorlinks=true,linkcolor=PrismTeal,urlcolor=PrismTeal,citecolor=PrismTeal}
\titleformat{\section}
  {\sffamily\bfseries\color{PrismNavy}\large}
  {\colorbox{PrismTeal}{\color{white}\strut\thesection}}{0.55em}{}
\titleformat{name=\section,numberless}
  {\sffamily\bfseries\color{PrismNavy}\large}
  {}{0pt}{}
\titlespacing*{\section}{0pt}{1.35ex plus .4ex minus .2ex}{0.65ex}
\titleformat{\subsection}
  {\sffamily\bfseries\color{PrismTeal}\normalsize}
  {\thesubsection}{0.45em}{}
\titlespacing*{\subsection}{0pt}{1.15ex plus .3ex minus .2ex}{0.45ex}

\DeclareCaptionFont{prismteal}{\color{PrismTeal}}
\setlist[itemize]{leftmargin=1.25em,itemsep=2pt,topsep=3pt,parsep=0pt}
\setlist[enumerate]{leftmargin=1.4em,itemsep=2pt,topsep=3pt,parsep=0pt}

\fancypagestyle{plain}{%
  \fancyhf{}%
  \fancyhead[L]{}%
  \fancyhead[R]{}%
  \fancyfoot[C]{\sffamily\footnotesize\color{PrismNavy}\thepage}%
}

\makeatletter
\renewenvironment{thebibliography}[1]
  {\section*{\refname}%
   \small\sloppy
   \list{[\arabic{enumiv}]}%
        {\settowidth\labelwidth{[#1]}%
         \leftmargin\labelwidth
         \advance\leftmargin\labelsep
         \usecounter{enumiv}%
         \let\p@enumiv\@empty
         }%
   \setlength{\itemsep}{2.5pt}%
   \setlength{\parsep}{0pt}%
   \setlength{\parskip}{0pt}}
  {\endlist}
\makeatother

\begin{document}
\twocolumn[
\begin{minipage}{\textwidth}
  \centering
  {\sffamily\bfseries\color{PrismNavy}\fontsize{21}{24}\selectfont
  PRISM-DR: Per-lesion Retinal Inference with Specialist Models for Diabetic Retinopathy\par}
  \vspace{7pt}
  {\color{PrismTeal}\rule{\textwidth}{1.2pt}}\par
  \vspace{7pt}
  {\sffamily\large Z\"ubeyr \"Ozeren\textsuperscript{a,b,*}\quad Tansel Uyar\textsuperscript{a}\par}
  \vspace{3pt}
  {\small
  \textsuperscript{a}Department of Biomedical Engineering, Baskent University, Ankara, T\"urkiye\par
  \textsuperscript{b}Department of Biomedical Engineering, Ankara University, Ankara, T\"urkiye\par}
  \vspace{3pt}
  {\footnotesize\textsuperscript{*}Corresponding author:
  \href{mailto:zubeyrsaidozeren@ankara.edu.tr}{zubeyrsaidozeren@ankara.edu.tr}\par
  \href{https://orcid.org/0009-0001-9606-4783}{ORCID 0009-0001-9606-4783}
  \quad\textbullet\quad
  \href{https://orcid.org/0000-0001-8083-2920}{ORCID 0000-0001-8083-2920}\par}
  \vspace{9pt}
  \noindent\colorbox{PrismPale}{%
    \parbox{0.955\textwidth}{%
      \vspace{4pt}\small
      {\sffamily\bfseries\color{PrismTeal}Abstract}\par\vspace{3pt}
      Diabetic retinopathy is a leading cause of preventable blindness; its
early lesions are small, low contrast, and easily missed in manual
screening. Most automated detectors handle the four non-proliferative DR
lesions: microaneurysms, hemorrhages, hard exudates, and soft exudates,
with a single multi-class model, even though these lesions differ
sharply in size, color, morphology, and prevalence, so a shared model
favors common, easy classes over rare, difficult ones. We present
PRISM-DR, a lesion-specific pipeline that trains one single-class
detector per lesion, each with its own configuration. From a raw fundus
image, the pipeline applies region of interest cropping, fundus-specific
preprocessing, four parallel YOLO detectors, tiling, per-lesion
ensembling of five cross-validation folds, and an inter-lesion
suppression step that resolves overlaps by physical lesion size and
clinical priority rather than confidence. Per lesion, the best of five
YOLO generations is selected, and augmentation is tuned by Bayesian
optimization. Trained on IDRiD with stratified five-fold
cross-validation, the system reaches a test mAP50 of 0.527 and F1 of
0.529, highest AP50 on hard exudates with 0.561. Without fine-tuning,
the models transfer well where the imaging scale is close to IDRiD and
degrade as field of view and resolution depart. These modest absolute
results reflect a small single-source training set and a difficult task;
however, treating each lesion as a separate detection problem is a
practical alternative to a single multi-class model.      \vspace{4pt}
    }%
  }\par
  \vspace{6pt}
  \begin{minipage}{0.97\textwidth}
    \footnotesize\sffamily
    \textbf{\color{PrismTeal}Keywords}\enspace Diabetic retinopathy; Lesion detection; Object
detection; YOLO; Color fundus images  \end{minipage}
\end{minipage}
\vspace{13pt}
]

\section{Introduction}

Diabetes mellitus is a chronic metabolic disease in which blood glucose stays persistently high, either because the pancreas produces too little insulin or because cells cannot use it effectively {[}1,2{]}. The 11th edition of the IDF Diabetes Atlas estimated that about 589 million adults between ages 20 and 79 were living with diabetes in 2024, roughly 11.1\% of that age group, and projected a rise to 853 million by the year 2050 {[}3{]}. Sustained hyperglycemia damages small vessels throughout the body {[}4{]}, and the eye is among the organs most often affected, with diabetic retinopathy (DR) being the most common of these microvascular complications {[}5{]}.

DR develops as chronic high glucose weakens the walls of the retinal capillaries, increasing vascular permeability, causing occlusions, and eventually driving abnormal new vessel growth {[}6,7{]}. It is one of the leading causes of preventable blindness in working-age adults {[}8{]}. The disease is usually asymptomatic in its early stages, and vision loss often appears only once it has advanced {[}9{]}. Large trials such as the Diabetic Retinopathy Study and the Early Treatment Diabetic Retinopathy Study showed that timely treatment can substantially reduce the risk of severe vision loss {[}10,11{]}, which makes regular screening essential. In many low- and middle-income countries, the shortage of trained ophthalmologists and limited infrastructure keep screening programs from reaching everyone who needs them {[}12{]}, so reliable automated tools have become a practical necessity.

Color fundus imaging is the most widely used modality for DR screening since it is non-invasive, inexpensive and broadly available {[}13,14{]}. Four primary non-proliferative lesions appear on these fundus images: microaneurysms (MA), hemorrhages (HE), hard exudates (EX) and soft exudates (SE). Their presence, number and distribution are the basis of the International Clinical Diabetic Retinopathy (ICDR) severity scale and of the treatment decisions that follow {[}15{]}. Microaneurysms are the earliest clinical sign, appearing as small dark-red dots usually under 125 micrometers across; dot hemorrhages can be almost the same size, while blot and flame hemorrhages are larger {[}16,17{]}. Hard exudates are sharply bordered yellow-white lipid deposits, and soft exudates are paler, fuzzy cotton-wool patches that signal retinal ischemia {[}16{]}. Detecting all four accurately and at the same time is what an ICDR-consistent early assessment requires.

Analyzing fundus images is still mostly manual, and it is open to several problems. Different graders interpret the same image differently, and this inter-observer variability lowers the reliability of a diagnosis {[}18,19{]}, especially for small lesions such as microaneurysms where experts often disagree {[}18{]}. Lesions that span only a few pixels are easy to miss, yet they are exactly the earliest and most useful findings for early intervention {[}20{]}. Detailed reading is also slow, and as the number of patients grows, long hours at the screen may lead to fatigue and reduced accuracy {[}21{]}.

These pressures have driven a large body of work on AI-based eye diagnosis in recent years {[}22--25{]}. However, systems of this kind mostly classify the DR stage rather than localize individual lesions. Lesion-level detection is the natural next step: it tells the clinician where the disease appears and in what form, which is the information the ICDR scale is built on. Classical machine learning methods relied on hand-crafted feature sets and struggled with the wide variability of lesion appearance {[}26,27{]}. Deep learning has largely removed that limitation, but it brings its own difficulties that the automated DR analysis literature has not fully addressed {[}28,29{]}.

Several gaps stand out. Many object detection studies target only one or two lesion types rather than all four {[}30--32{]}. Those that cover all four, on the other hand, train a single model for the whole task {[}33,34{]}, even though the lesions differ sharply in size, color, morphology and prevalence; a shared model tends to learn the easy lesions and neglect the hard ones, particularly MA {[}29{]}. When a high resolution fundus image is downscaled to a typical detector input of 512 or 640 pixels, a microaneurysm of a few pixels effectively disappears {[}29{]}. Preprocessing is rarely studied on its own; most work applies a fixed pipeline without measuring the actual effect on individual lesions {[}33,34{]}. The visual similarity between lesion types, such as dot hemorrhages and microaneurysms, can make different models trigger detections on the same region, yet few systems offer a principled way to resolve those overlaps, instead relying on in-model methods. Augmentation parameters are usually left at defaults, even though some transforms make little clinical sense on fundus images {[}35{]}.

This paper addresses those gaps with PRISM-DR (Per-lesion Retinal Inference with Specialist Models for Diabetic Retinopathy), an end-to-end system for the simultaneous detection of all four primary DR lesions. The main idea is lesion specificity: instead of one detector for everything, each lesion gets its own model and its own set of design choices. The main contributions are as follows:

\begin{itemize}
\item
  A lesion-specific pipeline in which the architecture, tiling strategy, augmentation policy, ensemble method and confidence threshold are selected separately for each lesion, optimized to its size, morphology and sample count,
\item
  A controlled, per-lesion comparison of five YOLO generations under identical data, preprocessing and hyperparameters, so that architectural differences can be read in isolation,
\item
  Per-lesion augmentation tuning by Bayesian optimization, with the search space shaped by the optical and clinical properties of fundus images,
\item
  An inter-lesion suppression step that resolves overlaps between different lesion models using physical and clinical knowledge rather than cross-model confidence,
\item
  A controlled preprocessing ablation and a zero fine-tuning external validation on three datasets, namely DDR, e-ophtha and TJDR, including an image-matching strategy for cross-dataset use.
\end{itemize}

\section{Related Work}

Public DR datasets vary widely in size and in the type of annotation they provide. Some offer only image-level DR grades, while others give pixel-level lesion masks or bounding boxes. Messidor and Messidor-2 {[}36,37{]}, EyePACS {[}38{]} and APTOS2019 {[}39{]} supply image-level grades and support classification tasks. While e-ophtha {[}40{]}, FGADR {[}41{]} and TJDR {[}42{]} provide only pixel-level annotations, DDR {[}29{]} and IDRiD {[}43{]} support both classification and segmentation tasks. Among these, datasets with bounding box annotations are scarce; only DDR provides those, which is one reason most detection studies derive boxes from segmentation masks. The ones with the pixel-level annotation masks are relevant to the proposed methodology.

Segmentation has been the more common route to lesion-level analysis. Furtado compared FCN and DeepLabv3 variants on IDRiD and used modified losses to improve microaneurysm IoU (Intersection over Union) {[}44{]}. Xu et al. added multi-scale feature fusion and channel attention to a U-Net {[}45{]}, and Shaukat et al. combined an Xception encoder with DeepLabv3 for joint segmentation and grading {[}46{]}. More recent networks include LezioSeg {[}47{]}, MLNet {[}48{]}, SDC-Net {[}49{]} and an attention-augmented DeepLab {[}50{]}. These methods offer precise lesion boundaries, but they do not directly report the count or location of individual lesions; extracting that requires extra post-processing such as connected-component analysis. Moreover, because segmentation operates at the pixel level, the models are computationally heavy {[}51{]}, and severe class imbalance often produces high false positive or false negative rates on small lesions {[}44{]}.

Object detection localizes each lesion with a bounding box and returns count, position, class and confidence scores in one step, which suits clinical decision support. Li et al. benchmarked Faster R-CNN, SSD and YOLOv3 when introducing DDR and reported very low detection scores, which underlined how hard the task is {[}29{]}. Alyoubi et al. combined YOLOv3 with a classifier on DDR {[}34{]}. Several YOLO-based studies followed: Santos et al. paired YOLOv5 with tiling on DDR and IDRiD {[}33{]}, Gao et al. tuned a YOLOv4 variant for microaneurysms only {[}31{]}, Zhang et al. detected microaneurysms in fluorescein angiography with MA-YOLO {[}52{]}, Pereira et al. combined YOLOR-CSP with Slicing Aided Hyper Inference (SAHI) {[}53{]}, Rizzieri et al. compared YOLOv8 and YOLOv9 on a small Messidor subset {[}54{]}, Han et al. proposed a lightweight YOLOv5 for small lesions {[}35{]}, and Hu et al. applied attention-augmented YOLOv8 to ultra-widefield images {[}55{]}.

Two further gaps motivate this work. First, the YOLO studies above each use different datasets, preprocessing and hyperparameters; thus, the effect of moving between YOLO generations on DR lesions cannot be isolated. Second, most of the studies do not validate their results on external datasets, validate on only private ones or rely on manually created labels; all of which make the reported results hard to interpret in terms of generalization and reproducibility. PRISM-DR tackles both with a controlled generation comparison and an explicit image-matching mode.

\section{Materials and Methods}

PRISM-DR takes a raw color fundus image and produces bounding-box detections for the four lesions. The image is region of interest (ROI) cropped and preprocessed once, then routed to four parallel lesion pipelines, each holding the architecture, tiling setting and augmentation policy chosen for its lesion and applying five cross-validation folds merged by a lesion-specific ensembling method. The four merged pools are then reconciled by an inter-lesion suppression step that resolves overlapping predictions between lesion types and provides the final output. Fig. 1 shows the overall architecture, and the following subsections describe each component of it.

\begin{figure*}[tbp]
\centering
\includegraphics[width=0.72\textwidth,height=0.85\textheight,keepaspectratio]{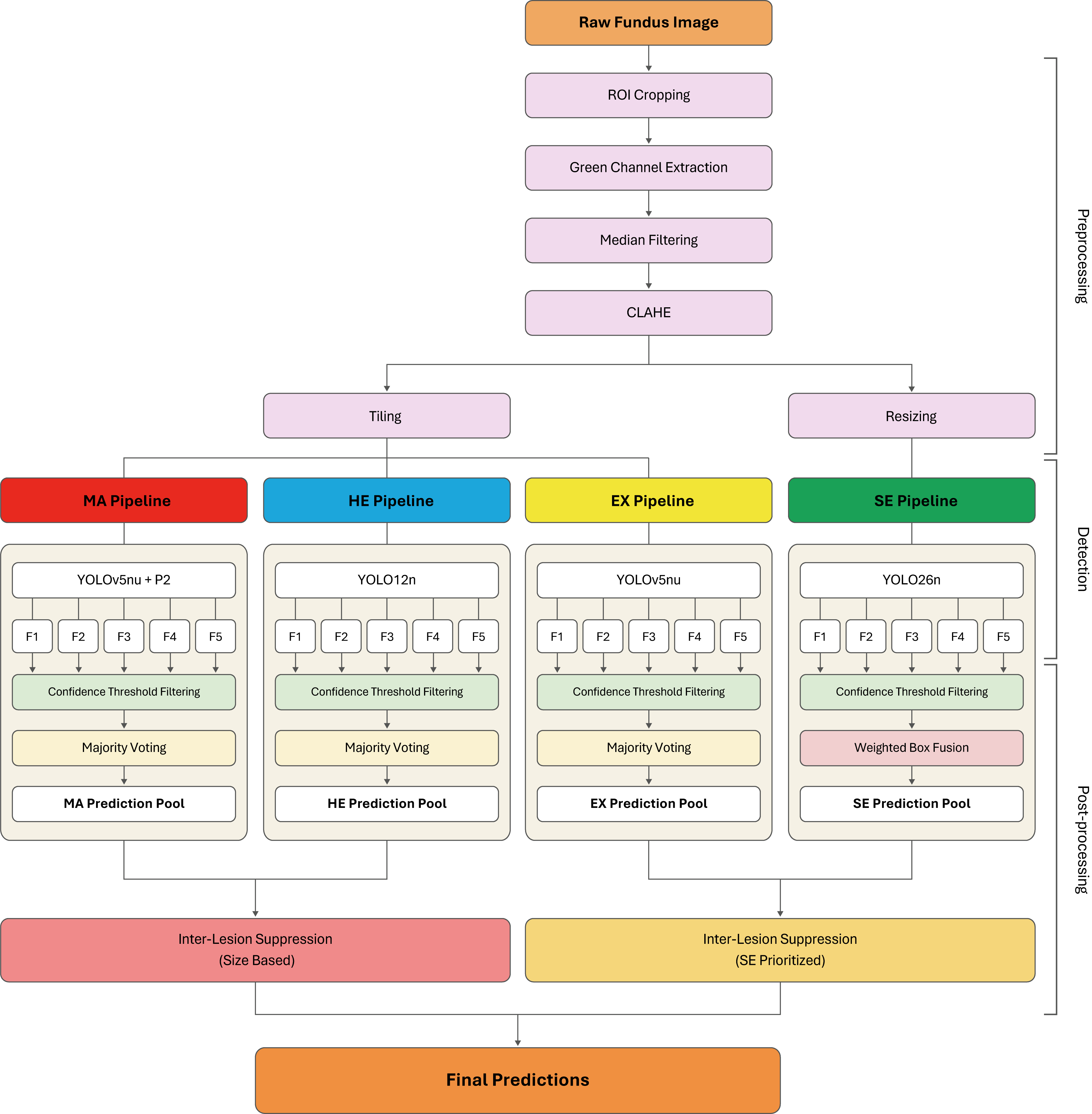}
\caption{Overall architecture of PRISM-DR. A raw color fundus image is ROI cropped and preprocessed once (green channel extraction, median filtering, CLAHE), then routed to four parallel lesion pipelines: microaneurysm (MA), hemorrhage (HE), hard exudate (EX), and soft exudate (SE). Each pipeline runs its own architecture and five cross-validation fold models (F1--F5), applies per-fold confidence-threshold filtering, and merges the folds with a lesion-specific rule (Majority Voting for MA/HE/EX; Weighted Boxes Fusion for SE). The four merged pools are reconciled by two inter-lesion suppression rules, size-based for the MA/HE pair, SE-priority for the EX/SE pair, to give the final predictions. MA, HE, and EX are tiled; SE is detected on the resized whole image.}
\label{fig:1}
\end{figure*}

\subsection{Datasets}

IDRiD {[}43{]} was the primary training and evaluation dataset. Its segmentation subset contains 81 high resolution images with 4288$\times$2848 pixels, and 50-degree field of view captured with a single Kowa VX-10$\alpha$ camera, each annotated at pixel level for the four lesions, and split by the authors into 54 training and 27 test images. A single camera and a fixed field of view keep the pixel scale constant and simplify preprocessing, but that also limits how well a model trained on IDRiD alone generalizes, which is the motivation behind externally validating the models on three additional datasets. e-ophtha {[}40{]} provides MA and EX annotations only; DDR {[}29{]} was collected from 147 hospitals and offers wide device and population diversity; and TJDR {[}42{]} includes a 133-degree ultra-widefield camera alongside a conventional one, so the same lesion can occupy different pixel sizes in the images. Table 1 summarizes the datasets used in this study.

A recurring property across all four datasets is severe class imbalance: lesion pixels cover only a small fraction of each image (about 2\% for IDRiD) {[}56{]}, and soft exudates are far rarer than all the other lesions. Fig. 2 shows an example image with its masks and boundaries.

\begin{table*}[tbp]
\centering
\caption{Datasets used in this study. IDRiD is the primary training and evaluation set; e-ophtha, DDR, and TJDR are held-out external validation sets. Columns give the publication year, number of images, native pixel resolution, field of view (FOV), the lesion types annotated, microaneurysms (MA), hemorrhages (HE), hard exudates (EX), soft exudates (SE), and the role in this study. For the external datasets the image count refers to the lesion-labeled subset only; resolution ranges (e.g., 1440$\times$960--2544$\times$1696) reflect mixed cameras within a dataset.}
\label{tab:1}
\footnotesize
\renewcommand{\arraystretch}{1.18}
\begin{tabular*}{\textwidth}{@{\extracolsep{\fill}}l c c c c c c@{}}
\toprule
\textbf{Dataset} & \textbf{Year} & \textbf{Images} & \textbf{Resolution (px)} & \textbf{FOV} & \textbf{Lesions} & \textbf{Use} \\
\midrule
IDRiD & 2018 & 81 & 4288$\times$2848 & 50\textdegree{} & MA, HE, EX, SE & Training + test \\
e-ophtha & 2013 & 463 & 1440$\times$960 -- 2544$\times$1696 & 45\textdegree{} & MA, EX & External \\
DDR & 2019 & 757 & 1380$\times$1382 -- 5184$\times$3456 & 45\textdegree{} & MA, HE, EX, SE & External \\
TJDR & 2023 & 561 & 2048$\times$2048 / 3912$\times$3912 & 45\textdegree{} / 133\textdegree{} & MA, HE, EX, SE & External \\
\bottomrule
\end{tabular*}
\end{table*}

\begin{figure*}[t]
\centering
\includegraphics[width=\textwidth,height=0.85\textheight,keepaspectratio]{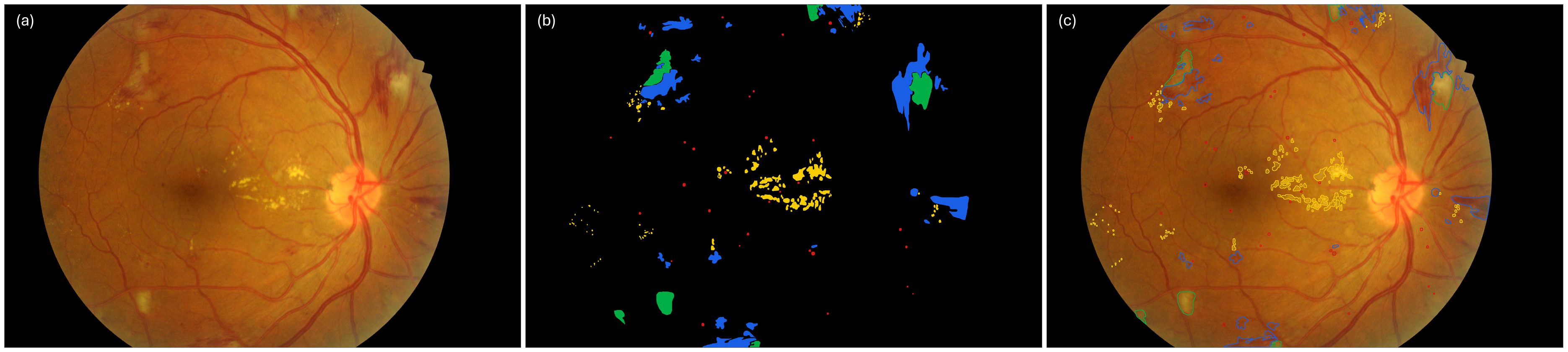}
\caption{Example fundus image from the IDRiD dataset with expert pixel-level annotations. (a) Original color fundus image; (b) lesion masks; (c) mask boundaries overlaid on the original image. In (b) and (c), red = microaneurysms (MA), blue = hemorrhages (HE), yellow = hard exudates (EX), green = soft exudates (SE). The four lesions differ markedly in size, color, and abundance; the observation that motivates the per-lesion design.}
\label{fig:2}
\end{figure*}

The segmentation datasets provide binary masks, not the bounding boxes an object detector requires, and one mask can hold many lesion instances. For each lesion the mask was loaded in grayscale and binarized, and the OpenCV connected-components method was applied with 8-connectivity, so each separate group of pixels received a unique label; the minimum and maximum coordinates of each component gave the box corners, which were normalized to the YOLO label format. Because the masks were drawn by expert ophthalmologists, no lower or upper size threshold was applied and every separate lesion was kept as one object, which keeps training and evaluation consistent with the expert annotations. Fig. 3 illustrates the conversion and Table 2 reports the final box counts.

\begin{table}[H]
\centering
\caption{Number of lesion bounding boxes per dataset after converting segmentation masks to boxes. Counts are listed separately for MA, HE, EX, and SE, with IDRiD split into its training and test partitions. A dash denotes a lesion not annotated in that dataset (e-ophtha labels only MA and EX). For DDR, e-ophtha, and TJDR the counts refer to the whole segmentation-labeled set.}
\label{tab:2}
\scriptsize
\renewcommand{\arraystretch}{1.18}
\begin{tabularx}{\columnwidth}{@{}l*{4}{Y}@{}}
\toprule
\textbf{Dataset} & \textbf{MA} & \textbf{HE} & \textbf{EX} & \textbf{SE} \\
\midrule
IDRiD (training) & 2412 & 1365 & 7526 & 112 \\
IDRiD (test) & 1085 & 535 & 3816 & 38 \\
DDR & 10380 & 12538 & 23659 & 1293 \\
e-ophtha & 1306 & -- & 2278 & -- \\
TJDR & 1507 & 3204 & 3352 & 503 \\
\bottomrule
\end{tabularx}
\end{table}

\begin{figure*}[t]
\centering
\includegraphics[width=\textwidth,height=0.85\textheight,keepaspectratio]{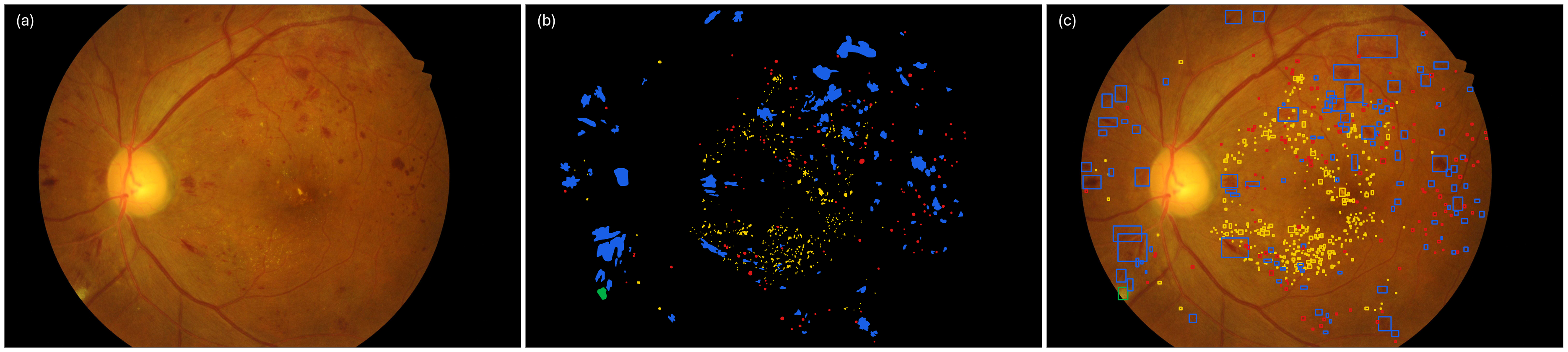}
\caption{Conversion of pixel-level segmentation masks to bounding boxes. (a) Original fundus image; (b) connected components extracted from the masks (OpenCV, 8-connectivity), each treated as one lesion instance; (c) the resulting bounding boxes overlaid on the image. Colors denote lesion type as in Fig. 2 (red = MA, blue = HE, yellow = EX, green = SE). Every separate lesion is kept as one box with no size threshold, matching the expert annotations.}
\label{fig:3}
\end{figure*}

\subsection{Experimental Setup}

Training and validation were performed on an NVIDIA A100 GPU via Google Colab, whereas visualization, inference, and external validation were carried out on a local workstation equipped with an NVIDIA RTX 4070 GPU and a Ryzen 5 7600X CPU. All random seeds were fixed to ensure reproducibility. The pipeline was implemented in Python 3.13 with PyTorch 2.10 {[}57{]}, and all five YOLO generations were trained and evaluated within a single codebase using Ultralytics 8.4.8 {[}58{]} under a common set of default settings, enabling a fair comparison across architectures. Image preprocessing was handled with OpenCV {[}59{]} and NumPy {[}60{]}, the data augmentation was tuned with Optuna {[}61{]}, and the dataset splits were generated using scikit-learn {[}62{]}.

Every model started from the COCO-pretrained weights that Ultralytics ships with each generation. Three settings were moved off the library defaults. The epoch budget was raised from 100 to 1000 while the early-stopping patience stayed at 100. The input size was evaluated as an experiment using 640, 960 and 1280 pixels. The cap on detections per image was raised from 300 to 10,000, because an advanced case can carry well over 300 lesions in a single fundus image, where the default would silently discard true detections. The rest was left as the library sets it. The batch size was 16, and the optimizer was left at auto, which lets Ultralytics choose one from the estimated iteration count and all models trained with AdamW. The initial and final learning rates, momentum and warmup epochs are left at defaults as 0.01, 0.01, 0.937 and 3 respectively. The random seed was fixed at 0 in every run.

Detections are matched to ground truth at an IoU of 0.5, and results are reported as AP50 per lesion, mAP50 over the four lesions, and F1 at each fold's own F1-optimal confidence threshold. Component experiments run on a stratified five-fold cross-validation of the 54 IDRiD training images, with the 27-image test set held out until the final system is evaluated. On the test set the per-fold thresholds were applied directly, with no threshold search on the test data. The metric definitions, the fold construction, and the threshold selection procedure are given in the Supplementary Materials.

\subsection{Image Preprocessing}

\begin{figure}[tbp]
\centering
\includegraphics[width=\columnwidth,height=0.85\textheight,keepaspectratio]{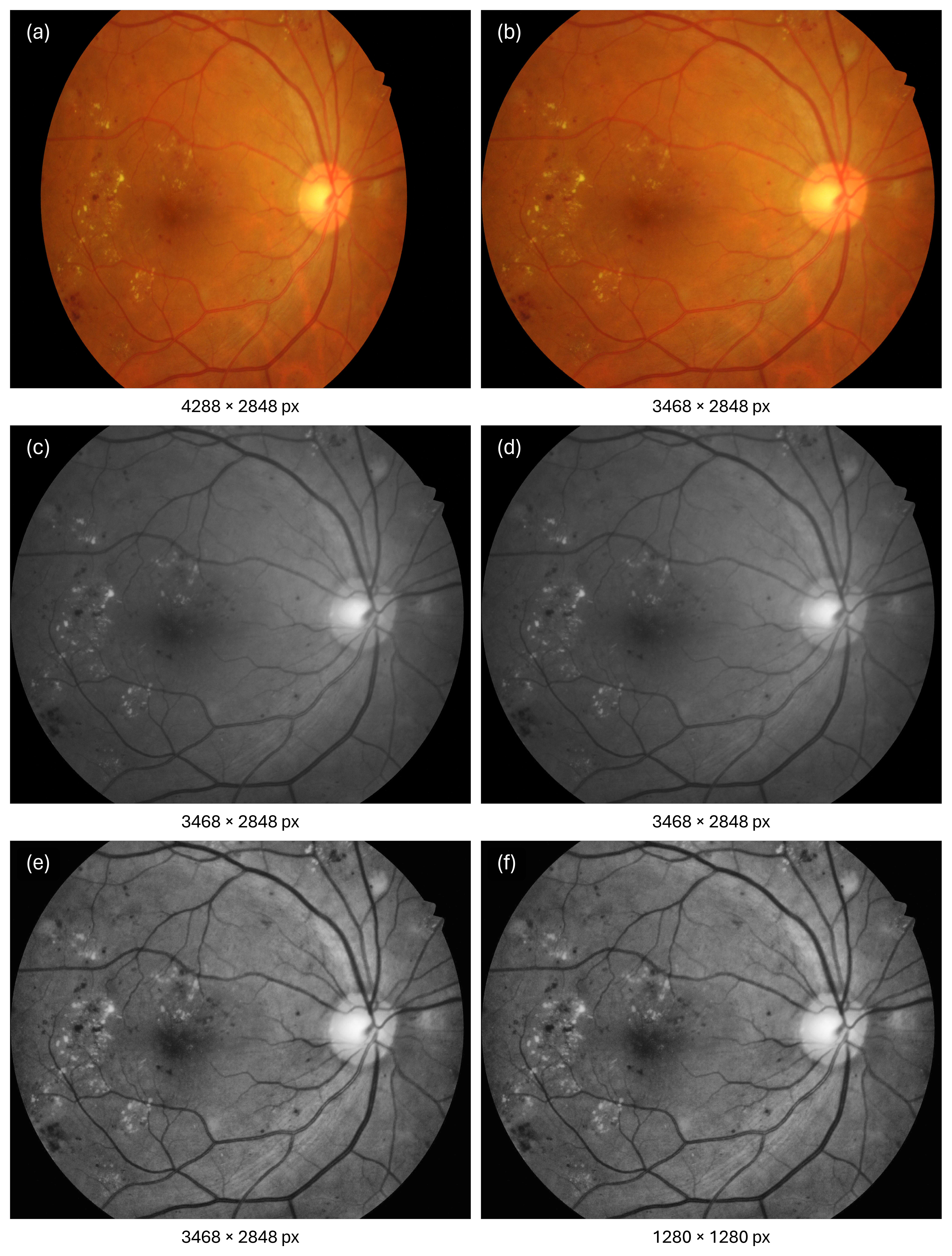}
\caption{The preprocessing pipeline applied to each fundus image, shown step by step: (a) original image; (b) image cropped to the region of interest (ROI), removing the black border; (c) green channel extraction, which gives the highest lesion contrast; (d) 5$\times$5 median filtering for noise suppression; (e) contrast-limited adaptive histogram equalization (CLAHE) for local contrast enhancement; (f) resizing to the detector input size (Lanczos interpolation). Pixel dimensions are shown beneath each panel.}
\label{fig:4}
\end{figure}

Fundus images usually carry a wide black border around the retinal region. Processing that border wastes computation and shrinks the retina after resizing. A small YOLOv8n detector was trained to localize the retinal region. ROI coordinates are computed once per image and cached so that every later step shares the same coordinate system. After ROI cropping the image passes through green channel extraction, median filtering, contrast limited adaptive histogram equalization (CLAHE) and resizing. The green channel gives the highest contrast since hemoglobin absorbs strongly at those wavelengths, so vessels and hemorrhages appear dark and exudates appear bright. A 5$\times$5 median filter then suppresses noise while preserving lesion edges; the kernel size was chosen using the lesion size distribution, since larger kernels would erase microaneurysms. CLAHE corrects the uneven illumination and enhances the contrast further. Resizing to the detector input size uses Lanczos interpolation, which handles both downscaling and the upscaling needed later for external validation. Fig. 4 shows the steps.

\subsection{Tiling}

\begin{figure*}[tbp]
\centering
\includegraphics[width=\textwidth,height=0.85\textheight,keepaspectratio]{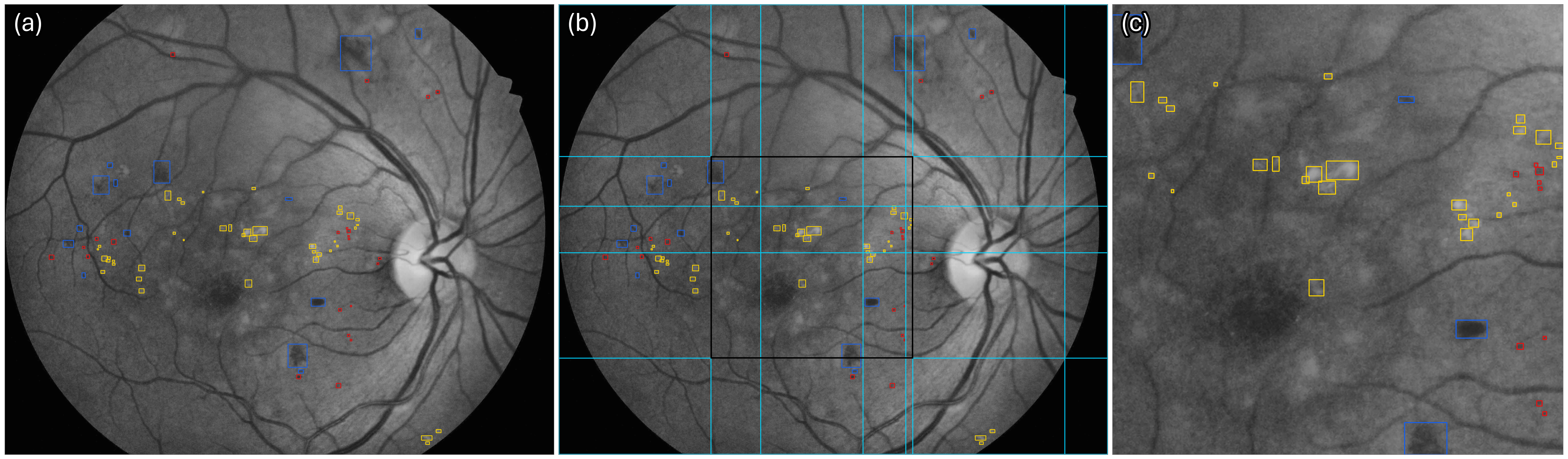}
\caption{The tiling procedure used for the MA, HE, and EX pipelines. (a) Preprocessed image with ground-truth lesion boxes (colors as in Fig. 2); (b) division into overlapping tiles; neighboring tiles overlap by 25\% so a lesion at a tile boundary appears whole in at least one tile; (c) a single extracted tile, fed to the detector at native resolution. A ground-truth box is retained in a tile only when at least 25\% of its area falls inside that tile. Soft exudates (SE) are the exception and are detected on the resized whole image without tiling.}
\label{fig:5}
\end{figure*}

Detectors take a fixed, relatively small input, and feeding a full fundus image means downscaling it. Downscaling a 4288$\times$2848 image to a 640-pixel input shrinks a 20$\times$20-pixel microaneurysm to roughly 3$\times$3 pixels. With a standard detection head whose finest stride is 8 pixels, an object that small is barely represented in the feature map and is, in practice, invisible to the model. Tiling resolves this: after preprocessing, the image is divided into tiles (patches) whose size equals the model input, so each tile is fed at native resolution with no downscaling at all, and small lesions are represented in the network weights at their true size.

Tile size parameter trades off a specific risk: smaller tiles preserve scale but cut the spatial context around a lesion and multiply the number of tiles, while larger tiles keep more context at the price of some effective resolution. Thus, tile size was treated as an experimental factor (640, 960 and 1280 pixels). Neighboring tiles overlap by 25\%, which guarantees that a lesion located at a tile boundary still appears whole in at least one tile. A ground-truth box is kept in a tile only when at least 25\% of its area falls inside that tile. This rule keeps the label when most of a boundary lesion is present but drops thin slivers that carry too little of the lesion to be learnable. Edge tiles are zero-padded to match the input size. Both lesion-bearing and lesion-free tiles are kept in training, so that the model sees healthy retina as well and produces fewer false positives at inference. Fig. 5 shows the tiling process.

Soft exudates are the deliberate exception and are detected on the resized image without tiling. SEs are large, low contrast patches that are known to be confused with the optic disc and bright artifacts {[}63{]}; thus, keeping these confusing factors in the same image helps the detection when model converges.

Each tile's detections are mapped back to global image coordinates using that tile's offset, and detections that land in the overlap between neighboring tiles are merged, so a lesion caught in two tiles is not counted twice.

\subsection{YOLO Generation Selection}

To isolate the effect of the architecture, five YOLO generations were compared for each lesion under identical conditions: YOLOv5u, YOLOv8, YOLO11, YOLO12, and YOLO26, all through one Ultralytics version. Only the nano variant of each generation was selected and used for two reasons: the 54-image training split makes larger variants prone to overfitting, and comparing the same variant keeps any difference attributable to the architecture rather than to model scale. With five generations, four lesions, and five folds, this comes to 100 model trainings, and the generation with the best validation mAP50 was selected per lesion.

For the microaneurysm model only, a P2 detection head was added. A YOLO detector predicts from a feature pyramid whose standard heads are P3, P4, and P5, which operate at strides 8, 16, and 32 pixels respectively, so even the finest standard head places one prediction cell every 8 pixels. A microaneurysm is often only a few pixels across, so it can fall entirely inside a single P3 cell and be localized too coarsely to match its ground-truth box. The P2 head adds a finer level at stride 4, adding this head doubles the spatial density of predictions and gives a small lesion enough representation in the feature maps.

\subsection{Per-lesion Augmentation Optimization}

Augmentation parameters were tuned separately for each lesion with Bayesian optimization using Optuna and its Tree-Structured Parzen Estimator (TPE) {[}64{]}, over the ten parameters in Table 3, with overall validation mAP50 as the objective. The search space was shaped by the physics and clinical properties of fundus imaging. Hue shift was fixed at zero, since lesion color carries clinical meaning (red for MA and HE, yellow-white for EX and SE) and shifting it would teach unrealistic color distributions. Perspective and shear were left disabled since a fundus camera records on a flat plane at a fixed angle. The lower bound of horizontal flip was set to 0.5, rather than 0, because left and right eyes are mirror images, so a horizontal flip is anatomically valid and a useful free source of variety. The search ran in two stages: 40 trials per fold of 50 epochs each; then the top three configurations per fold were re-evaluated across all five folds, and the candidate with the highest mAP50 was selected for that lesion.

\begin{table}[H]
\centering
\caption{Search space for the per-lesion data-augmentation optimization. The table lists the ten tuned Ultralytics parameters with their ranges and sampling type (step size for continuous parameters; 'discrete' or 'categorical' otherwise). The optimization objective was overall validation mAP50. Hue shift, perspective, and shear were fixed at zero (disabled) on optical and clinical grounds and were not part of the search.}
\label{tab:3}
\scriptsize
\renewcommand{\arraystretch}{1.18}
\begin{tabularx}{\columnwidth}{@{}l c c >{\raggedright\arraybackslash}X@{}}
\toprule
\textbf{Parameter} & \textbf{Range} & \textbf{Step / type} & \textbf{Description} \\
\midrule
hsv\_s & 0.0 -- 1.0 & 0.1 & Saturation shift \\
hsv\_v & 0.0 -- 0.5 & 0.1 & Brightness shift \\
degrees & 0 -- 30 & 5 & Rotation angle \\
translate & 0.0 -- 1.0 & 0.1 & Translation fraction \\
scale & 0.0 -- 1.0 & 0.1 & Scale gain \\
flipud & 0.0 / 0.5 / 1.0 & discrete & Vertical flip probability \\
fliplr & 0.5 / 1.0 & discrete & Horizontal flip probability \\
mosaic & 0 / 1 & categorical & Four image mosaic \\
mixup & 0 / 1 & categorical & Image blending \\
cutmix & 0 / 1 & categorical & Region cut and paste \\
\bottomrule
\end{tabularx}
\end{table}

\subsection{Lesion-specific Ensembling}

Each lesion's five-fold models give complementary predictions, and using a single fold directly is unstable, because a fold is trained on only four fifths of an already small dataset and its validation set is a random draw. The fold models were therefore merged with an ensemble. A standard non-maximum suppression merge was rejected due to its attribute of discarding the agreement information across folds, which is the very signal that separates a reliable detection from a single fold's noise. The choice between the two agreement-based methods was then matched to each lesion's fold-to-fold behavior.

For MA, HE and EX, majority voting keeps a prediction only when at least three folds place a box on it at an IoU of 0.5. These lesion models generate many detections, including noise-driven false positives from vessel crossings and bright artifacts, and requiring agreement filters out the boxes a single fold invents, which raises precision. SE, by contrast, has by far the fewest training examples and the highest fold-to-fold variance, so even a 3/5 rule would throw away many true detections seen by only one or two folds. Weighted Boxes Fusion (WBF) {[}65{]} is used for it instead: rather than voting boxes in or out, it keeps low-agreement detections but scales their confidence by the fraction of folds that contributed. Also, it averages the box coordinates weighted by confidence, which preserves true positives while sharpening localization where the fold models slightly disagree.

\subsection{Inter-lesion Suppression}

\begin{figure*}[tbp]
\centering
\includegraphics[width=0.72\textwidth,height=0.85\textheight,keepaspectratio]{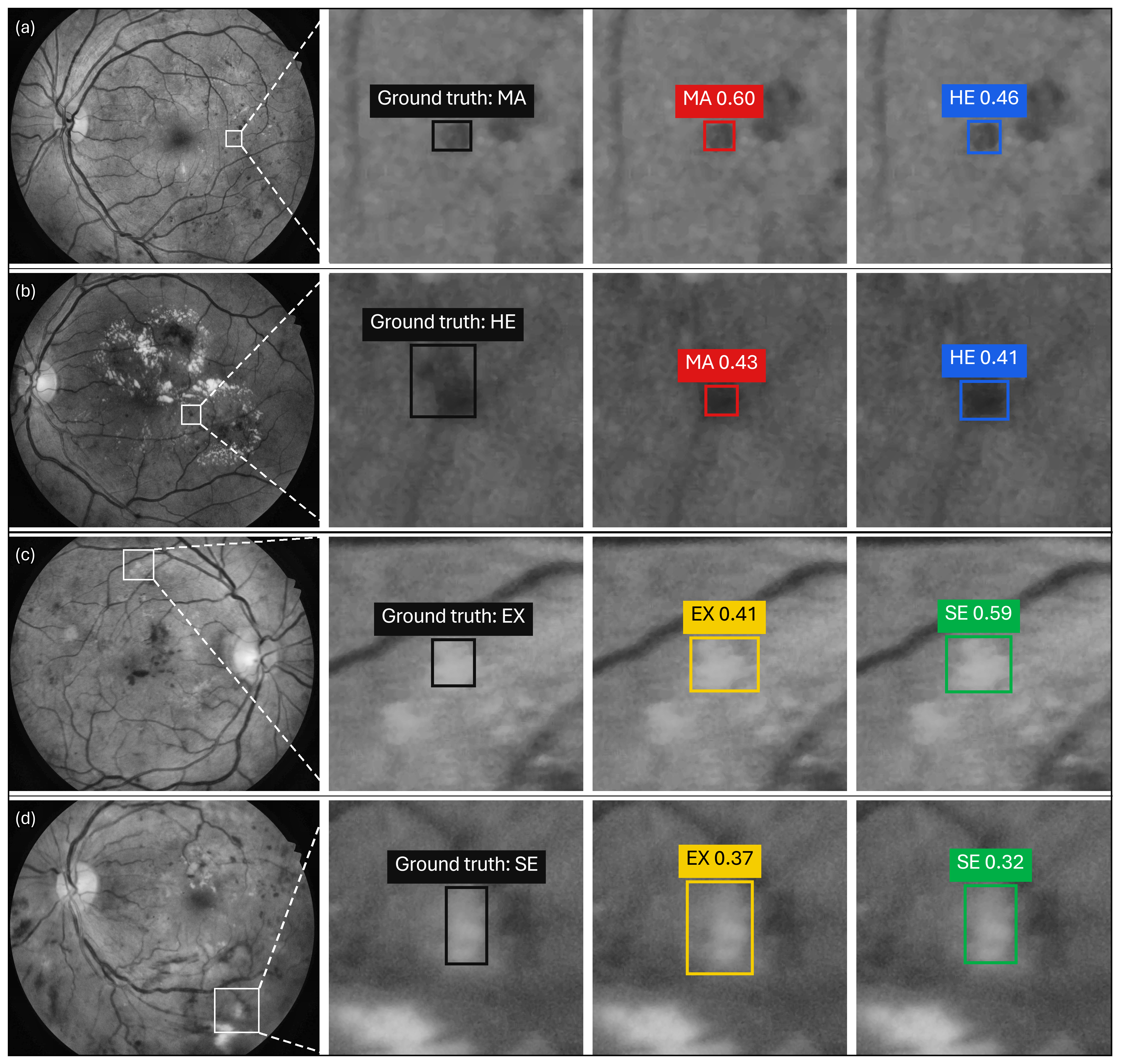}
\caption{Examples of overlapping predictions from the independently trained per-lesion detectors, resolved by inter-lesion suppression. Each row (a--d) shows one case in which two competing single-class models both fire on the same location. (a) A true microaneurysm (MA) detected by both the MA and hemorrhage (HE) models; (b) a true hemorrhage detected by both the MA and HE models: the red lesion pair, resolved by physical lesion size. (c) A true hard exudate (EX) detected by both the EX and soft exudate (SE) models; (d) a true soft exudate detected by both the EX and SE models: the bright lesion pair, resolved in favor of SE. In each row, from left to right: the preprocessed fundus image with the region of interest cropped; the zoomed region with its ground-truth lesion label; and the box predicted by each competing model, each shown in its own panel with its confidence score.}
\label{fig:6}
\end{figure*}

Since the lesion models are independent, two of them can place boxes on the same region, for instance an MA and an HE model on a dot hemorrhage, or an EX and an SE model on a bright patch. A naive fix for this is to keep the higher-confidence box, but confidence scores from independently trained models are not directly comparable, so a confidence rule might suppress true predictions. We instead use inter-lesion suppression, a confidence-free rule defined for two clinically grounded pairs. For the red lesion pair (MA and HE), overlapping boxes are resolved by size: the detection is labeled MA when the average box size is below a threshold of 38 pixels and HE otherwise. The threshold was set in two complementary ways that agree with each other: it follows from the clinical upper bound of microaneurysm diameter of 125 micrometers converted to pixels using the Gullstrand schematic eye model {[}66,67{]}, and it matches the empirical box-size distributions of MA and HE in the IDRiD training set, where most MA boxes fall below it and most HE boxes above it. For the bright lesion pair (EX and SE), overlaps are always resolved in favor of SE, and the overlapping EX box is suppressed. This is justified by data, since the abundant EX model otherwise dominates and lowers SE recall, and by clinical grounds, since the soft exudate marks nerve-fiber ischemia and represents a more advanced finding. Both rules rest on quantities that do not depend on the model or training setup. Fig. 6 shows example overlaps of both lesion groups.

\subsection{External Validation Inference}

\begin{figure}[tbp]
\centering
\includegraphics[width=\columnwidth,height=0.85\textheight,keepaspectratio]{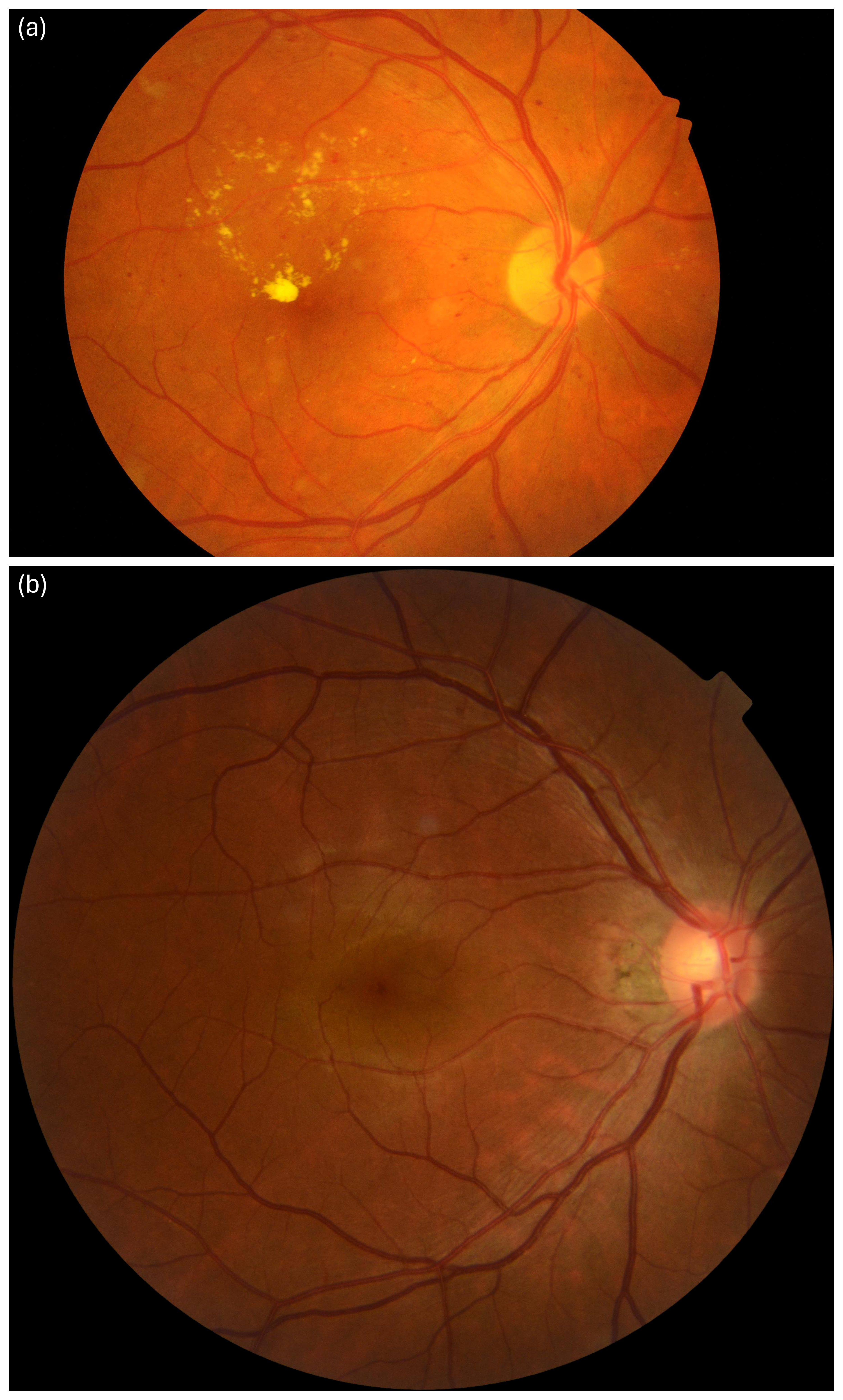}
\caption{The two fundus-image framings handled by the IDRiD-matching inference mode. (a) A wider, landscape 'cut-off' framing whose top and bottom are cropped by the sensor (aspect ratio $\geq$ 1.2), rescaled to 3400$\times$2848; (b) a full, near-circular retina (aspect ratio < 1.2), rescaled to 3400$\times$3400.}
\label{fig:7}
\end{figure}

The IDRiD-trained models were applied to DDR, e-ophtha, and TJDR with no fine-tuning and unchanged confidence thresholds. The difficulty is that these datasets differ from IDRiD in terms of resolution and field of view, so a lesion covers different number of pixels than the models learned to expect. To reduce this gap, each image is ROI cropped and then rescaled so that its retinal region approaches the pixel scale of IDRiD. Fundus images appear in two common framings: a full, near-circular retina with an aspect ratio close to one, and a frame whose top and bottom are cut off by the sensor, giving a wider, landscape image like IDRiD's. The aspect ratio after cropping selects one of two targets whose own aspect ratio is close to the image, which keeps the rescaling close to isotropic, so lesion shapes are preserved: images with an aspect ratio of 1.2 or above are rescaled to 3400$\times$2848, and the rest to 3400$\times$3400. The 1.2 cutoff is the value that separates these two framings, and both targets approximate the pixel extent of the IDRiD retinal region after ROI cropping, so a lesion lands at roughly the scale the models trained on. Preserving aspect ratio matters clinically, since a near-circular microaneurysm rescaled with unequal axes would look elliptical and no longer match the trained circular appearance. After that, the images follow the preprocessing and 1280-pixel tiling as in training, and SE is run on a single 1280-pixel input without tiling. Fig. 7 showcases the difference between two framings.

\section{Results}

Results use mAP50 as the primary metric and macro F1 score as the secondary metric, with per-fold F1-optimal thresholds. Component studies are reported on the stratified five-fold cross-validation as mean $\pm$ standard deviation across folds; the final system and the external validation are reported on held-out data. Up to the model selection stage, the YOLO26n model served as the baseline detector, since it was the most recent and official generation when this study began.

\subsection{Preprocessing and Input Size}

The preprocessing pipeline was evaluated as a whole, comparing it against no preprocessing at two input sizes, with a single four-class model and ROI cropping kept on throughout, results can be found in Table 4. Its effect was strongly lesion dependent. At 640 pixels, adding the green channel, median filter, and CLAHE raised MA from 0.239 to 0.323 AP50 and EX from 0.290 to 0.333, while HE barely moved and SE was essentially flat. The pattern follows the lesions' visibility. MA and EX are small and low in contrast, so the green channel in which hemoglobin-rich and lipid-rich structures stand out most, together with CLAHE's local contrast boost, makes them easier to separate from the background. Most HE and SE are already large and salient, so the same steps add little. Input size had the larger overall effect: moving from 640 to 1280 pixels lifted every lesion, because at 640 a few pixel lesion is too small to represent and more pixels recover it directly. At 1280 the extra preprocessing still helped, but by less, since the higher resolution already supplies the small lesions with enough representation. We therefore adopted full preprocessing at 1280 input size for the rest of the study. That the four lesions responded differently to the same preprocessing and input size changes is the first concrete evidence that a single shared configuration is suboptimal, and it motivates the per-lesion design.

\begin{table*}[t]
\centering
\caption{Effect of preprocessing and input size, evaluated with a single four-class model and ROI cropping kept on throughout. Values are per-lesion AP50 (average precision at an intersection-over-union, IoU, of 0.5) for MA, HE, EX, and SE, and their mean over the four lesions (mAP50, 'All'), reported as mean $\pm$ standard deviation over the five cross-validation folds. Configurations combine input size (640\textsuperscript{2} vs 1280\textsuperscript{2} pixels) with and without the preprocessing pipeline (green channel, median filter, CLAHE). Bold marks the best value in each column.}
\label{tab:4}
\footnotesize
\renewcommand{\arraystretch}{1.18}
\begin{tabularx}{\textwidth}{@{}l*{5}{Y}@{}}
\toprule
\textbf{Configuration} & \textbf{MA} & \textbf{HE} & \textbf{EX} & \textbf{SE} & \textbf{All} \\
\midrule
640\textsuperscript{2}, no preprocessing & 0.239 $\pm$ 0.027 & 0.344 $\pm$ 0.028 & 0.290 $\pm$ 0.045 & 0.532 $\pm$ 0.072 & 0.351 $\pm$ 0.043 \\
640\textsuperscript{2}, with preprocessing & 0.323 $\pm$ 0.031 & 0.346 $\pm$ 0.032 & 0.333 $\pm$ 0.044 & 0.531 $\pm$ 0.036 & 0.383 $\pm$ 0.036 \\
1280\textsuperscript{2}, no preprocessing & 0.387 $\pm$ 0.037 & 0.417 $\pm$ 0.024 & 0.431 $\pm$ 0.038 & 0.604 $\pm$ 0.088 & 0.460 $\pm$ 0.047 \\
1280\textsuperscript{2}, with preprocessing & \textbf{0.395 $\pm$ 0.026} & \textbf{0.424 $\pm$ 0.027} & \textbf{0.439 $\pm$ 0.034} & \textbf{0.612 $\pm$ 0.128} & \textbf{0.468 $\pm$ 0.054} \\
\bottomrule
\end{tabularx}
\end{table*}

\subsection{Per-lesion and Single-model Detection}

Keeping the 1280-pixel preprocessed setup, four independent single-class models were compared against the single four-class model under identical training (Table 5). The per-lesion design improved AP50 on all four lesions and raised overall mAP50 from 0.468 to 0.495 and F1 from 0.461 to 0.491. The gains were uneven and they show exactly where a shared model fails: MA improved most, then EX, HE, and SE least.

A four-class detector shares one backbone, one set of feature maps, and one set of weights across lesions that look different, from a few-pixel red microaneurysm to a large, diffuse cotton-wool spot. The network must compress all of them into a single representation, and it tends to settle on whatever serves the easy, abundant lesions, leaving the others poorly modeled. A dedicated model, on the other hand, has no such conflict: it can shape its features, anchor scales, and operating point around one lesion's texture, brightness and size. Giving each lesion its own model, and its own optimal confidence threshold, removes the competition and lets every lesion be learned more successfully.

The per-lesion ordering of the gains follows directly from these mechanisms. MA gains most because it is the lesion most penalized by sharing: a small, low contrast object is easily suppressed when the same weights must also fit the abundant EX and the large SE, and a dedicated MA model can spend its capacity and anchor scales on tiny structures. SE gains least, but for an encouraging reason. With the largest average area, it is already the best-learned lesion even under the shared model, so there is little headroom left to recover.

Beyond the immediate accuracy gain, the per-lesion split is what makes the rest of the pipeline possible. Because each lesion now has its own model, it can also have its own architecture, tiling decision, augmentation policy, and ensemble rule, each chosen on its own merits in the following sections, rather than forced into a single compromise across four very different targets. The remaining experiments quantify those per-lesion choices.

\begin{table*}[t]
\centering
\caption{Single four-class model versus four independent single-class (per-lesion) models, trained under identical settings at 1280\textsuperscript{2} input with full preprocessing. Values are per-lesion AP50 for MA, HE, EX, and SE and their mean, as mean $\pm$ standard deviation over the five folds; the 'Change' row is the per-lesion difference.}
\label{tab:5}
\footnotesize
\renewcommand{\arraystretch}{1.18}
\begin{tabularx}{\textwidth}{@{}l*{5}{Y}@{}}
\toprule
\textbf{Configuration} & \textbf{MA} & \textbf{HE} & \textbf{EX} & \textbf{SE} & \textbf{All} \\
\midrule
Single model & 0.395 $\pm$ 0.026 & 0.424 $\pm$ 0.027 & 0.439 $\pm$ 0.034 & 0.612 $\pm$ 0.128 & 0.468 $\pm$ 0.054 \\
Per-lesion models & 0.453 $\pm$ 0.026 & 0.445 $\pm$ 0.028 & 0.463 $\pm$ 0.031 & 0.617 $\pm$ 0.086 & 0.495 $\pm$ 0.043 \\
Change & +0.058 & +0.021 & +0.024 & +0.005 & +0.027 \\
\bottomrule
\end{tabularx}
\end{table*}

\subsection{Tiling}

Three tile sizes were compared using the per-lesion models, against whole-image inference (Table 6). For MA, HE, and EX, AP50 peaked at 1280 pixels. A larger tile carries more surrounding context, which helps the model separate a microaneurysm from a vessel crossing or an exudate from a bright artifact, and it splits fewer lesions across tile boundaries, so fewer fall below the 25\% area threshold that decides whether a box is kept in a tile. The 25\% overlap between neighboring tiles serves the same purpose, giving every lesion a chance to appear whole in at least one tile rather than being cut in two at a seam.

SE behaved in the opposite way and was best without tiling (0.617 AP50, against 0.511 at 1280). Two effects work against tiling here. Soft exudates are the largest lesions, so one can span more than one tile and be cut into fragments that each fall under the retention threshold, costing labels. Additionally, their faint appearance is easy to confuse with the optic disc or with bright artifacts, and on the full image the model can lean on global cues, the position of the disc and macula and the overall lesion layout, to tell them apart, whereas a single tile strips that context away. The final system therefore tiles MA, HE, and EX at 1280 pixels and detects SE on the resized image.

\begin{table*}[t]
\centering
\caption{Per-lesion AP50 as a function of tile size, using the per-lesion models, with the mean over the four lesions; values are mean $\pm$ standard deviation over the five folds. Rows compare three tile sizes (640\textsuperscript{2}, 960\textsuperscript{2}, 1280\textsuperscript{2} pixels) against whole-image inference.}
\label{tab:6}
\footnotesize
\renewcommand{\arraystretch}{1.18}
\begin{tabularx}{\textwidth}{@{}l*{5}{Y}@{}}
\toprule
\textbf{Configuration} & \textbf{MA} & \textbf{HE} & \textbf{EX} & \textbf{SE} & \textbf{All} \\
\midrule
640\textsuperscript{2} tiles & 0.445 $\pm$ 0.033 & 0.414 $\pm$ 0.028 & 0.510 $\pm$ 0.025 & 0.373 $\pm$ 0.073 & 0.436 $\pm$ 0.040 \\
960\textsuperscript{2} tiles & 0.457 $\pm$ 0.029 & 0.454 $\pm$ 0.014 & 0.497 $\pm$ 0.031 & 0.542 $\pm$ 0.115 & 0.488 $\pm$ 0.047 \\
1280\textsuperscript{2} tiles & \textbf{0.476 $\pm$ 0.046} & \textbf{0.475 $\pm$ 0.020} & \textbf{0.527 $\pm$ 0.030} & 0.511 $\pm$ 0.075 & 0.497 $\pm$ 0.043 \\
No tiling & 0.453 $\pm$ 0.026 & 0.445 $\pm$ 0.028 & 0.463 $\pm$ 0.031 & \textbf{0.617 $\pm$ 0.086} & 0.495 $\pm$ 0.043 \\
\bottomrule
\end{tabularx}
\end{table*}

\subsection{Generation Selection}

Table 7 reports per-lesion AP50 for the five generations under identical settings. The best generation differs by lesion: YOLOv5nu for MA (0.502) and EX (0.529), YOLO12n for HE (0.486), and YOLO26n for SE (0.617). MA and SE had clear winners, while HE and EX showed small gaps between generations. The pattern argues against a general rule that the newest or largest architecture always wins; the right choice depends on lesion size, appearance, and sample count. Adding a P2 head to the MA model raised its AP50 from 0.502 to 0.526, consistent with the higher-resolution feature map better representing smaller lesions. Relative to the YOLO26n baseline, the selected per-lesion architectures raised overall mAP50 from 0.523 to 0.539, with MA gaining 0.050 AP50.

\begin{table*}[t]
\centering
\caption{Per-lesion AP50 for the nano variant of five YOLO generations, each trained under identical data, preprocessing, and hyperparameters so that differences reflect architecture rather than model scale. Values are mean $\pm$ standard deviation over the five folds.}
\label{tab:7}
\footnotesize
\renewcommand{\arraystretch}{1.18}
\begin{tabularx}{\textwidth}{@{}l*{4}{Y}@{}}
\toprule
\textbf{Generation} & \textbf{MA} & \textbf{HE} & \textbf{EX} & \textbf{SE} \\
\midrule
YOLOv5nu & \textbf{0.502 $\pm$ 0.031} & 0.481 $\pm$ 0.020 & \textbf{0.529 $\pm$ 0.038} & 0.595 $\pm$ 0.080 \\
YOLOv8n & 0.484 $\pm$ 0.035 & 0.476 $\pm$ 0.017 & 0.523 $\pm$ 0.038 & 0.606 $\pm$ 0.082 \\
YOLO11n & 0.474 $\pm$ 0.038 & 0.483 $\pm$ 0.020 & 0.521 $\pm$ 0.041 & 0.581 $\pm$ 0.077 \\
YOLO12n & 0.488 $\pm$ 0.037 & \textbf{0.486 $\pm$ 0.034} & 0.522 $\pm$ 0.050 & 0.570 $\pm$ 0.080 \\
YOLO26n & 0.476 $\pm$ 0.046 & 0.475 $\pm$ 0.020 & 0.527 $\pm$ 0.030 & \textbf{0.617 $\pm$ 0.086} \\
\bottomrule
\end{tabularx}
\end{table*}

\subsection{Per-lesion Augmentation}

The optimized configurations differ noticeably by lesion. MA turned off every geometric transform (rotation, translation, vertical flip) and all three mixing methods. A microaneurysm is only a few pixels across, so a small rotation or shift can move it off the pixel grid or blur it beyond recognition, and the seams produced by mosaic, mixup, and cutmix introduce tiny artifacts that the model would mistake for real microaneurysms; the only augmentations kept were a mild saturation and brightness jitter and horizontal flip. EX enabled mosaic but disabled cutmix, with moderate geometric transforms: cutmix pastes rectangular patches whose hard straight edges resemble the sharp borders of a hard exudate, so it was suppressed, while mosaic, which rescales and tiles whole images, exposes the model to exudates at more scales without inventing false edges. HE took the widest morphology-driven variety, with the highest translation, because hemorrhages appear in several morphologies (dot, blot, and flame) that need wide variety to generalize. SE enabled all three mixing methods and the strongest transforms (degrees 15, scale 1.0): it is the rarest lesion, so augmentation diversity matters most for it, and its large, soft, low-frequency appearance tolerates mixing and scaling that would destroy a microaneurysm.

Against the default augmentation, tuning helped every lesion. The largest gain was on HE (+0.041 AP50), where the default settings seemingly underrepresented the morphological variety, with smaller but consistent gains on MA, EX, and SE.

\begin{table}[H]
\centering
\caption{Final selected data augmentation configuration for each lesion model, chosen by the per-lesion optimization of Table 3. Mosaic, mixup, and cutmix are on/off flags; hsv\_s and hsv\_v are HSV saturation/brightness gains; flipud and fliplr are flip probabilities; degrees, translate, and scale are geometric-transform magnitudes. Hue shift, perspective, and shear were disabled for all lesions.}
\label{tab:8}
\scriptsize
\renewcommand{\arraystretch}{1.18}
\begin{tabularx}{\columnwidth}{@{}l*{4}{Y}@{}}
\toprule
\textbf{Parameter} & \textbf{MA} & \textbf{HE} & \textbf{EX} & \textbf{SE} \\
\midrule
hsv\_s & 0.2 & 0.7 & 0.6 & 0.7 \\
hsv\_v & 0.3 & 0.3 & 0.1 & 0.5 \\
degrees & 0 & 10 & 5 & 15 \\
translate & 0.0 & 0.3 & 0.1 & 0.0 \\
scale & 0.8 & 0.6 & 0.6 & 1.0 \\
flipud & 0.0 & 0.5 & 0.5 & 0.5 \\
fliplr & 0.5 & 0.5 & 0.5 & 0.5 \\
mosaic & 0 & 1 & 1 & 1 \\
mixup & 0 & 0 & 0 & 1 \\
cutmix & 0 & 1 & 0 & 1 \\
\bottomrule
\end{tabularx}
\end{table}

\subsection{Ensembling and Inter-lesion Suppression}

After the augmentation, the final system is locked and tested on the test set. The lesion-specific ensemble improved every lesion over the average single-fold result, raising overall mAP50 from 0.487 to 0.511 and F1 from 0.491 to 0.511. HE gained most, since its morphological variety makes fold predictions inconsistent and voting filters that noise. Inter-lesion suppression then raised mAP50 to 0.527 and F1 to 0.529. Its per-lesion effect is concentrated on SE, which gained 0.052 AP50 because the SE-priority rule stops the abundant EX model from suppressing overlapping true SE; HE gained 0.010 AP50 from the size rule, which relabels high-confidence MA boxes that fall on dot hemorrhages; and EX changed by only 0.001, because the suppressed EX boxes were false positives that did not contribute to the EX score.

\begin{table}[H]
\centering
\caption{Per-lesion AP50 and overall mAP50 on the 27-image IDRiD test set across the final pipeline stages: the mean over the five fold models ('Per fold'), after lesion-specific ensembling ('+ Ensemble'), and after inter-lesion suppression ('+ Inter-lesion suppression', the final system). Columns give AP50 for MA, HE, EX, and SE and their mean.}
\label{tab:9}
\scriptsize
\renewcommand{\arraystretch}{1.18}
\begin{tabularx}{\columnwidth}{@{}l*{5}{Y}@{}}
\toprule
\textbf{Stage} & \textbf{MA} & \textbf{HE} & \textbf{EX} & \textbf{SE} & \textbf{All} \\
\midrule
Per fold (mean) & 0.530 & 0.440 & 0.556 & 0.424 & 0.487 \\
+ Ensemble & 0.551 & 0.475 & 0.560 & 0.458 & 0.511 \\
+ Inter-lesion suppression (final) & 0.554 & 0.485 & 0.561 & 0.510 & 0.527 \\
\bottomrule
\end{tabularx}
\end{table}

\subsection{Final System and Inference Speed}

Table 10 reports the final per-lesion performance after all stages. Overall mAP50 is 0.527 and F1 is 0.529. EX has the highest AP50 (0.561), followed by MA (0.554), SE (0.510), and HE (0.485). The strong MA result is notable, since the microaneurysm is the earliest sign of disease and the hardest lesion to detect, and it is the lesion that the per-lesion design helped most. Fig. 8 shows the confusion matrix and Fig. 9 shows qualitative detections for each lesion.

Tiling buys small lesion accuracy at the price of running many tiles per image, so inference speed had to be managed. On the 27-image IDRiD test set, the final system runs about 3 seconds per image. The baseline SAHI comparison and operation-wise time needed to process a single image are presented in the Supplementary Materials.

\begin{table}[H]
\centering
\caption{Final per-lesion performance of the complete PRISM-DR system on the 27-image IDRiD test set. Rows give precision, recall, F1, and AP50 for MA, HE, EX, and SE; the 'All' column is the mean over the four lesions. Precision, recall, and F1 are computed at each fold's F1-optimal confidence threshold, with detections matched to ground truth at IoU = 0.5.}
\label{tab:10}
\scriptsize
\renewcommand{\arraystretch}{1.18}
\begin{tabularx}{\columnwidth}{@{}l*{5}{Y}@{}}
\toprule
\textbf{Metric} & \textbf{MA} & \textbf{HE} & \textbf{EX} & \textbf{SE} & \textbf{All} \\
\midrule
Precision & 0.600 & 0.534 & 0.604 & 0.421 & 0.540 \\
Recall & 0.484 & 0.436 & 0.587 & 0.632 & 0.535 \\
F1 & 0.536 & 0.480 & 0.595 & 0.505 & 0.529 \\
AP50 & 0.554 & 0.485 & 0.561 & 0.510 & 0.527 \\
\bottomrule
\end{tabularx}
\end{table}

\begin{figure}[tbp]
\centering
\includegraphics[width=\columnwidth,height=0.85\textheight,keepaspectratio]{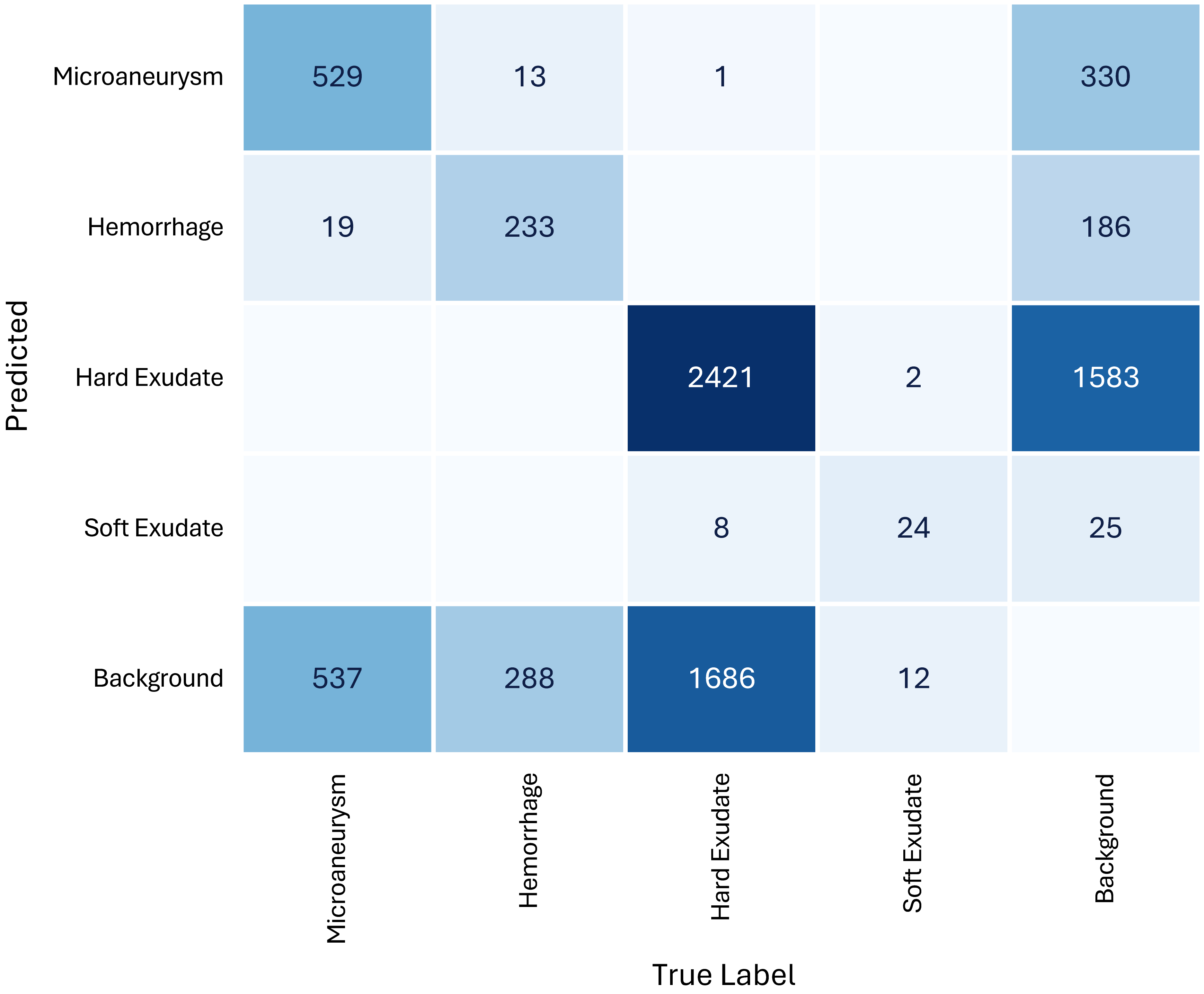}
\caption{Confusion matrix of the final system on the 27-image IDRiD test set, with detections matched to ground truth at IoU = 0.5. Rows are predicted classes and columns are true classes; MA, HE, EX, and SE are the four lesions, and 'Background' collects missed lesions (false negatives, in the bottom 'predicted-background' row) and spurious detections (false positives, in the right-hand 'true-background' column). Cell values are detection counts; off-diagonal cells within the four-lesion block are inter-lesion confusions.}
\label{fig:8}
\end{figure}

\begin{figure*}[tbp]
\centering
\includegraphics[width=\textwidth,height=0.85\textheight,keepaspectratio]{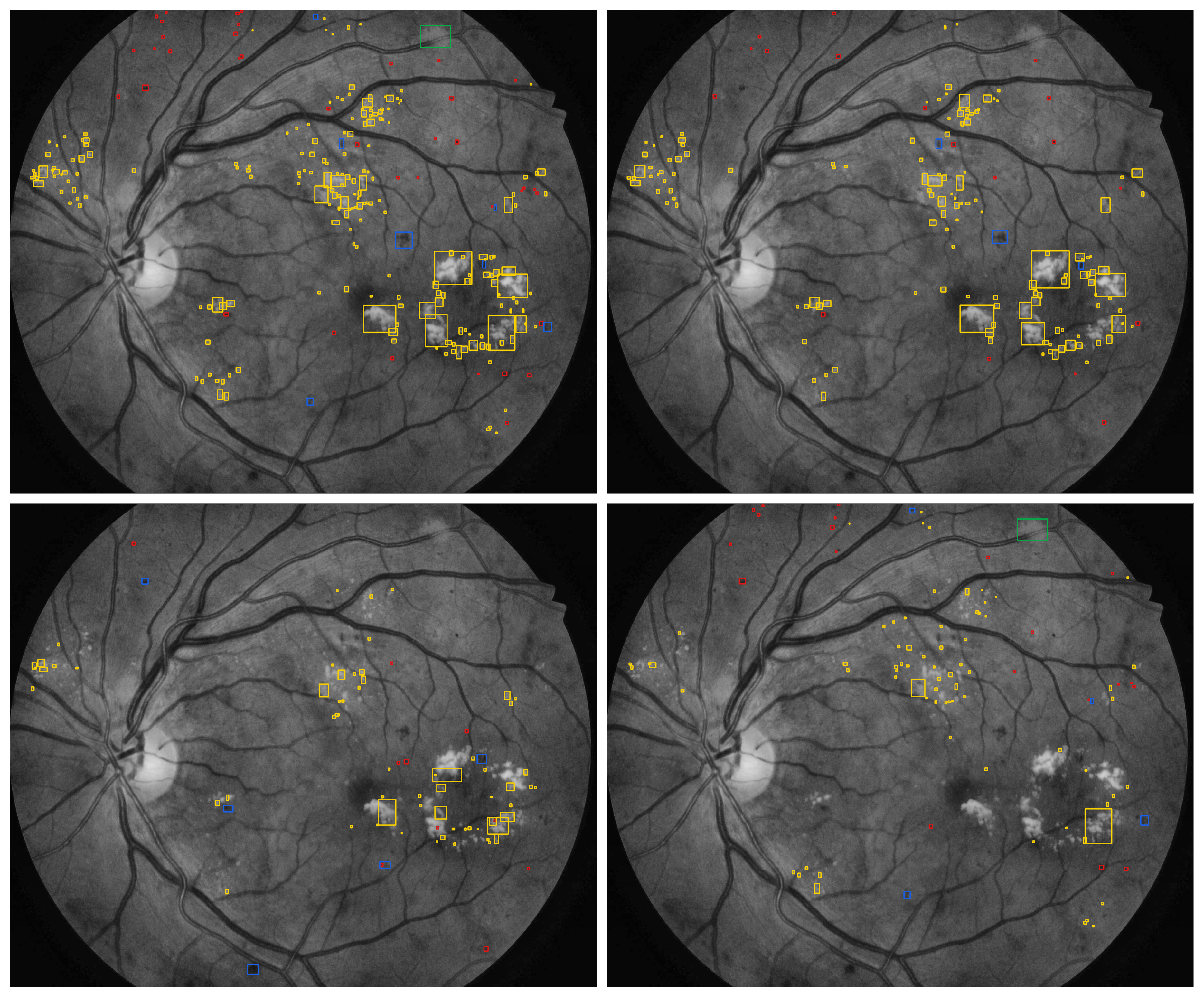}
\caption{Qualitative detection results on a representative IDRiD test image, with all four lesion types shown together. The four panels display the same preprocessed image: ground-truth boxes (top left), true positives: correct detections (top right), false positives: spurious detections (bottom left), and false negatives: missed lesions (bottom right). Predictions are the full system output, per-lesion five-fold ensemble with inter-lesion suppression, matched to ground truth at an IoU of 0.5. Box colors denote lesion type: red = microaneurysms (MA), blue = hemorrhages (HE), yellow = hard exudates (EX), green = soft exudates (SE).}
\label{fig:9}
\end{figure*}

\subsection{External Validation}

The IDRiD-trained models were applied to e-ophtha, DDR, and TJDR with no fine-tuning, using the IDRiD-matching inference mode. On e-ophtha, whose resolution is closest to IDRiD, MA reached 0.628 AP50, higher than the 0.554 the same model reaches in-domain on the IDRiD test set, and the overall score was 0.483 mAP50; the high MA number reflects both the matched scale and the relative homogeneity of e-ophtha. On DDR the overall score fell to 0.159 mAP50, which is consistent with images pooled from 147 hospitals differing widely in camera, quality, and annotation style; HE and EX transferred best. On TJDR the overall score was 0.267 mAP50, with HE again the most transferable lesion and MA among the weakest, because the 133-degree ultra-widefield images spread the retina over about three times the area of IDRiD and shrink an already tiny lesion below the trained scale. The drop from IDRiD to the external sets is expected for zero fine-tuning and is largest where the imaging geometry differs most.

\begin{table}[H]
\centering
\caption{Zero-fine-tuning external validation of the IDRiD-trained models on e-ophtha, DDR, and TJDR, applied through the IDRiD-matching inference mode (ROI crop and rescale to approximate IDRiD's pixel scale) with unchanged confidence thresholds. Values are per-lesion AP50 for MA, HE, EX, and SE and their mean; a dash denotes a lesion not labeled in that dataset.}
\label{tab:11}
\scriptsize
\renewcommand{\arraystretch}{1.18}
\begin{tabularx}{\columnwidth}{@{}l*{5}{Y}@{}}
\toprule
\textbf{Dataset} & \textbf{MA} & \textbf{HE} & \textbf{EX} & \textbf{SE} & \textbf{All} \\
\midrule
e-ophtha & 0.628 & -- & 0.339 & -- & 0.483 \\
DDR & 0.067 & 0.257 & 0.169 & 0.143 & 0.159 \\
TJDR & 0.195 & 0.427 & 0.185 & 0.261 & 0.267 \\
\bottomrule
\end{tabularx}
\end{table}

\subsection{Comparison with Prior Work}

Tables 12 and 13 place the system among detection studies on IDRiD, DDR, and e-ophtha. The referred studies differ in data splits, mask-to-box conversion, and tiling strategies, all of which shift the scores, so exact ranking would be misleading. On IDRiD, the system\textquotesingle s 0.527 mAP50 is in the range of the reported object detection results, with its highest per-lesion AP50 on EX and a strong MA result. For the external datasets, the more meaningful comparison is internal. On e-ophtha, the microaneurysm AP50 of 0.628, obtained with no fine-tuning, indicates that the MA detector transfers well once the imaging scale is matched, rather than that it surpasses any particular e-ophtha-trained model. On DDR, the 0.159 mAP50 falls within the range reported by models trained and tested on DDR, which is a reasonable outcome for a zero fine-tuning transfer to images pooled from 147 hospitals. TJDR has no published detection baseline, so the system can serve as an early reference point there.

\begin{table*}[tbp]
\centering
\caption{Comparison with published object detection studies on IDRiD. Values are per-lesion AP50 (MA, HE, EX, SE) and overall mAP50; a dash indicates a lesion not reported. 'Split' is the train:validation:test (or train:test) image partition, and 'Original' denotes the dataset's official split.}
\label{tab:12}
\footnotesize
\renewcommand{\arraystretch}{1.18}
\begin{tabularx}{\textwidth}{@{}ll*{5}{Y}@{}}
\toprule
\textbf{Method} & \textbf{Split} & \textbf{MA} & \textbf{HE} & \textbf{EX} & \textbf{SE} & \textbf{All} \\
\midrule
Mask R-CNN {[}68{]} & 50:20:30 & -- & -- & -- & -- & 0.346 \\
ResNet34+FPN+CBAM {[}69{]} & 80:20 & 0.644 & -- & -- & -- & 0.644 \\
YOLOR-CSP+SAHI {[}70{]} & 50:20:30 & 0.395 & 0.322 & 0.517 & 0.411 & 0.411 \\
PRISM-DR (proposed) & Original & 0.554 & 0.485 & 0.561 & 0.510 & 0.527 \\
\bottomrule
\end{tabularx}
\end{table*}

\begin{table*}[tbp]
\centering
\caption{Comparison with published object detection studies on DDR and e-ophtha. Values are per-lesion AP50 and overall mAP50; dashes indicate lesions not reported. The PRISM-DR rows are zero-fine-tuning external validation, whereas the comparison methods were trained on the target dataset.}
\label{tab:13}
\footnotesize
\renewcommand{\arraystretch}{1.18}
\begin{tabularx}{\textwidth}{@{}ll*{5}{Y}@{}}
\toprule
\textbf{Dataset} & \textbf{Method} & \textbf{MA} & \textbf{HE} & \textbf{EX} & \textbf{SE} & \textbf{All} \\
\midrule
DDR & YOLOv5s {[}71{]} & 0.003 & 0.059 & 0.034 & 0.100 & 0.051 \\
DDR & Mask R-CNN {[}68{]} & 0.104 & 0.158 & 0.252 & 0.155 & 0.167 \\
DDR & YOLOR-CSP+SAHI {[}70{]} & 0.138 & 0.218 & 0.329 & 0.203 & 0.222 \\
DDR & YOLOv5+ {[}35{]} & -- & -- & -- & -- & 0.352 \\
DDR & PRISM-DR (external) & 0.067 & 0.257 & 0.169 & 0.143 & 0.159 \\
\midrule
e-ophtha & Mask R-CNN {[}72{]} & -- & -- & -- & -- & 0.437 \\
e-ophtha & ResNet34+FPN+CBAM {[}69{]} & 0.622 & -- & -- & -- & 0.622 \\
e-ophtha & PRISM-DR (external) & 0.628 & -- & 0.339 & -- & 0.483 \\
\bottomrule
\end{tabularx}
\end{table*}

\FloatBarrier
\section{Discussion}

The per-lesion design began with an observation in our baseline results: among the four lesions, the microaneurysm was consistently the weakest under a single multi-class detector. We interpreted this as a limitation of the shared model setting rather than of the lesion itself. A single detector applies the same configuration to four lesions that differ in size, contrast, and frequency, and no single setting is optimal for all of them at once. Assigning each lesion its own model removes this constraint, because design decisions can then be made for one lesion rather than compromised across four. This improved detection accuracy on every lesion, and it opened a range of per-lesion choices that a shared model cannot accommodate: the architecture, the augmentation policy, the tiling decision, the ensembling method, and the confidence threshold can each be set for one lesion. This flexibility is what justifies the added cost of maintaining separate models.

One such choice was the detector architecture itself. Rather than assuming that the newest or largest model would perform best, we compared five YOLO generations for each lesion under identical conditions and selected the generation empirically. No single generation was best across all lesions, and a newer or larger model did not consistently yield better results. Optimizing augmentation separately for each lesion was a further consequence of the per-lesion design: with an independent model per lesion, augmentations could be adapted to each lesion\textquotesingle s appearance and clinical characteristics rather than fixed across all four.

Tiling was motivated on two grounds. The first was precedent: recent detection works apply tiling specifically to recover small lesions such as microaneurysms. The second was the interaction between lesion size and detector resolution. A YOLO detector represents objects down to its finest stride of eight pixels, so a lesion smaller than this in the network input cannot be localized reliably. Because fundus images are downscaled to a fixed input size, this imposes an effective size floor in the original image: at a 640-pixel input, lesions below approximately 48 pixels fall beneath the limit, and at 1280 pixels, those below approximately 24 pixels. These ranges encompass most microaneurysms and a portion of the smaller hemorrhages and hard exudates, which is why tiling, by presenting each region at native resolution, improved their detection.

The soft exudate was treated as an exception for two reasons. Its instances are consistently larger than this size floor, so it gains nothing from the resolution that tiling preserves. It also depends heavily on surrounding context, as its appearance is readily confused with the optic disc and with bright imaging artifacts, and that context is lost once an image is divided into tiles. Consistent with those, every tiling configuration reduced soft exudate performance relative to full-image inference.

Due to the per-lesion design choice, the risk of independent lesion models placing boxes on the same region appears. When this happens, the overlap has to be resolved. We chose not to base this decision on detection confidence. Scores produced by separately trained models are not directly comparable, so a confidence-based rule would be unreliable. For the microaneurysm and hemorrhage pair, which share a common clinical origin, the two lesions can be separated by size. This rule does not extend to the hard and soft exudate pair, for two reasons: the two lesions arise from different clinical processes, and they do not share a common size threshold. Overlaps in this pair were resolved instead by clinical priority.

Applied to the external datasets without fine-tuning, the system reached overall scores of 0.483 mAP50 on e-ophtha, 0.267 on TJDR, and 0.159 on DDR. The stronger result on e-ophtha follows from its closer match to IDRiD in imaging scale, whereas the lower scores on DDR and TJDR are the expected outcome of zero fine-tuning transfer to datasets that differ more widely in resolution and field of view.

The published detection studies differ in dataset splits, tiling strategies, and mask-to-box conversion, and these differences shift the reported scores enough that a direct ranking would be misleading; the contribution is therefore better read as a proof of concept for the per-lesion approach than as a performance benchmark. However, one methodological point is worth noting. When tiling is applied with overlapping tiles, a single lesion may be detected in more than one tile, and these duplicate detections must be merged before evaluation, as otherwise they inflate the reported scores. Established slicing frameworks such as SAHI {[}73{]} include this merging step, but custom tiling pipelines must implement it explicitly, and several tiling-based studies do not describe how it is handled.

Several limitations remain. The training set is small, with 54 images in the IDRiD split, which constrains model fitting and contributes to the fold-to-fold variance, largest for the soft exudates. Generalization also weakens on the external datasets, most sharply on DDR and TJDR, where the imaging conditions depart further from IDRiD. The study is also confined to the YOLO family; other object detection architectures were not investigated and may be better suited to particular lesions. Inference remains relatively slow, at roughly three seconds per image, which may be limiting for high-throughput screening. Finally, the reliance on segmentation masks converted to bounding boxes introduces an approximation of its own, and the scarcity of datasets with native bounding-box annotations restricts both training and evaluation.

\section{Conclusion}

We presented PRISM-DR, a lesion-specific pipeline that detects the four primary diabetic retinopathy lesions using a separate model for each rather than a single multi-class detector. From a raw color fundus image, the system performs ROI cropping and a fixed preprocessing sequence before routing the image to four parallel lesion pipelines. Each lesion is assigned to its own architecture, augmentation policy, tiling decision, ensembling method, and confidence threshold. Small lesions are detected on native-resolution tiles, while the soft exudates are detected on the whole image. Overlapping predictions between lesion models are reconciled by a confidence-free inter-lesion suppression step based on lesion size and clinical priority. The controlled experiments showed that the per-lesion design improved detection on all four lesions relative to a single multi-class model, and that the most effective architecture, tiling decision, and augmentation policy differed from one lesion to another.

The central conclusion is that treating each lesion as an independent detection problem is a suitable alternative to a single multi-class model, provided it is supported by an appropriate configuration and lesion-specific optimization. Beyond the accuracy it provides, the per-lesion formulation offers considerable flexibility, since each component of the pipeline can be adapted to the properties of an individual lesion rather than compromised across all four. The same property makes the framework straightforward to extend, as a new lesion can be added as an additional independent model without disturbing the others. The approach still requires further optimization, particularly in inference speed, but the results establish it as a viable proof of concept for multi-class lesion detection frameworks.

On the IDRiD test set, the system reached 0.527 mAP50 and 0.529 F1, with the strongest per-lesion results on hard exudates and microaneurysms. Applied to three external datasets without fine-tuning, it transferred well where the imaging scale was close to IDRiD and degraded as the field of view and resolution departed, with the highest overall score on e-ophtha and lower scores on TJDR and DDR.

Several directions follow from this work. The per-lesion principle can be extended to the stages that are still shared across lesions, including preprocessing, input resolution, and the loss function, and to the detector architecture itself. Preprocessing in particular could be specialized to each lesion, for example a green channel and CLAHE pipeline for the low contrast red lesions and a different channel selection or denoising setting for the bright ones. A larger, multi-source training set spanning additional cameras and fields of view would be expected to improve generalization. Extending the system from lesion-level detection toward an image- or patient-level output, such as an ICDR severity grade, would connect it more directly to the screening decision made in clinical practice. Finally, lighter models and further inference optimization would reduce the per-image processing time toward the requirements of high-throughput screening.

\section*{CRediT authorship contribution statement}
\addcontentsline{toc}{section}{CRediT authorship contribution statement}

\textbf{Z\"ubeyr \"Ozeren:} Conceptualization, Methodology, Software, Formal analysis, Investigation, Data curation, Visualization, Writing -- original draft. \textbf{Tansel Uyar:} Conceptualization, Methodology, Supervision, Writing -- review \& editing.

\section*{Declaration of competing interest}
\addcontentsline{toc}{section}{Declaration of competing interest}

The authors declare that they have no known competing financial interests or personal relationships that could have appeared to influence the work reported in this paper.

\section*{Funding}
\addcontentsline{toc}{section}{Funding}

This research did not receive any specific grant from funding agencies in the public, commercial, or not-for-profit sectors.

\section*{Ethics statement}
\addcontentsline{toc}{section}{Ethics statement}

This study was conducted using publicly available, fully de-identified retinal image datasets. No new data involving human participants or animals were collected, and no identifiable personal information was accessed. Therefore, ethical approval was not required.

\section*{Data availability}
\addcontentsline{toc}{section}{Data availability}

This study used only publicly available datasets. IDRiD, DDR, e-ophtha,
and TJDR can be accessed through their original publications and
repositories cited in this paper. The code implementation of the
pipeline is available at: \url{https://github.com/zubeyrozeren/PRISM-DR}.

\FloatBarrier
\balance

\clearpage
\FloatBarrier

\section*{Supplementary Materials}
\addcontentsline{toc}{section}{Supplementary Materials}

\setcounter{table}{0}
\renewcommand{\thetable}{S\arabic{table}}
\setcounter{equation}{0}
\renewcommand{\theequation}{S\arabic{equation}}

\subsection*{Evaluation Metrics and Cross Validation}

Detections are matched to ground-truth boxes by intersection over union, the overlap of two boxes divided by their union:
\begin{equation}
\mathrm{IoU}(A,B)=\frac{\mathrm{area}(A\cap B)}{\mathrm{area}(A\cup B)}
\end{equation}

A detection is considered as a true positive (TP) when it matches a ground-truth box at IoU $\geq$ 0.5, and otherwise a false positive (FP); an unmatched ground-truth box is a false negative (FN). Precision, recall, and their harmonic mean F1 score are calculated as follows:
\begin{equation}
\mathrm{Precision}=\frac{\mathrm{TP}}{\mathrm{TP}+\mathrm{FP}}
\end{equation}
\begin{equation}
\mathrm{Recall}=\frac{\mathrm{TP}}{\mathrm{TP}+\mathrm{FN}}
\end{equation}
\begin{equation}
\mathrm{F1\ Score}=\frac{2\cdot\mathrm{Precision}\cdot\mathrm{Recall}}{\mathrm{Precision}+\mathrm{Recall}}
\end{equation}

Average precision (AP) is the area under the precision-recall curve, computed with 101-point interpolation following the COCO convention, and AP50 denotes AP at the IoU threshold of 0.5. The primary metric, mean average precision (mAP50), is the mean of the per-lesion AP50 values over the $C$ number of lesion classes, calculated as:
\begin{equation}
\mathrm{AP}=\int_{0}^{1} p(r)\,dr
\end{equation}
\begin{equation}
\mathrm{mAP50}=\frac{1}{C}\sum_{c=1}^{C}\mathrm{AP50}_{c}
\end{equation}

With only 54 training images, a single split would make results depend on that specific split, so performance was evaluated with stratified five-fold cross-validation, about 43 training and 11 validation images per fold. Because the number of lesions per image varies widely, each image was assigned a stratum by its lesion density and distributed across folds by stratified sampling, so every fold represents all lesions comparably. The 27-image test set was excluded from cross-validation entirely and used only to measure the final system, which guarantees that the reported results are independent of model selection, hyperparameter tuning and threshold setting, and prevents any data leakage between development and test.

Detector performance depends on the confidence threshold, and a single fixed value is not optimal across folds whose confidence distributions differ. Each fold model's threshold was set on its own validation set by maximizing the F1 score, a balanced operating point between precision and recall, that is particularly important in medical applications. Each model first produced predictions at a low initial threshold of 0.001; precision, recall and F1 were then computed over a range of confidence values to form an F1-confidence curve at an IoU of 0.5, and the confidence at its maximum was stored.

\subsection*{Tiling and Batched Inference}

Although tiling helps small lesion detection, it also multiplies the computational cost per image. SAHI {[}73{]}, a commonly used sliced inference framework, processes the tiles of an image sequentially in its default mode. Each tile is a separate forward pass and for each tile it launches the model once more and moves data to the GPU. This sequential approach leaves the GPU underused. The routine used in this study changes how the tiles are executed. All tiles of an image are stacked into a single batch and passed through the model in one forward pass, which keeps the GPU saturated. The per-tile detections are then mapped back to global image coordinates using each tile's offset, and detections in the overlapping regions are merged so that a lesion captured in two tiles is not counted twice. The batched method returns the same detections as sequential slicing, only the order of execution differs.

Table S1 compares the standard sliced inference library SAHI, which sends tiles to the GPU one at a time, with the batched tiling routine. On the 27-image IDRiD test set, sequential SAHI took 14.64 s per image; the batched routine cut this to 3.08 s, a 4.75-fold speedup, by running all tiles of an image in a single GPU pass; and adding CPU prefetch, which reads and preprocesses the next image while the current one runs on the GPU, reached 3.04 s, a 4.82-fold speedup. Predictions were matched against SAHI to confirm that the speedup costs no accuracy.

\begin{table}[H]
\centering
\caption{Inference-speed comparison on the IDRiD test set (27 images), on the RTX 4070 / Ryzen 5 7600X machine.}
\label{tab:14}
\scriptsize
\renewcommand{\arraystretch}{1.18}
\begin{tabularx}{\columnwidth}{@{}l*{4}{Y}@{}}
\toprule
\textbf{Configuration} & \textbf{Total (s)} & \textbf{Per image (ms)} & \textbf{FPS} & \textbf{Speedup} \\
\midrule
Sequential SAHI & 395.35 & 14{,}642 & 0.07 & -- \\
Batched tiling & 83.18 & 3{,}080 & 0.32 & 4.75$\times$ \\
Batched tiling + CPU prefetch & 81.97 & 3{,}035 & 0.33 & 4.82$\times$ \\
\bottomrule
\end{tabularx}
\end{table}

Table S2 breaks the per-image time into operations. Tiled inference (about 2.07 s) and the tiling and resize step (about 0.93 s) dominate, while preprocessing (median filter and CLAHE, about 47 ms together) and the post-processing steps (confidence filtering, ensembling, and inter-lesion suppression, under 26 ms combined) are minor. The breakdown also explains why detecting SE on the whole image is cheap rather than costly: without tiling, the SE pipeline avoids the per-tile multiplication that drives the inference and tiling cost for the other three lesions, so handling SE differently improves its accuracy without adding to the dominant cost. End to end, the system processes an image in about 3.07 s on the test hardware.

\begin{table}[H]
\centering
\caption{Per-image processing time of the complete system on the IDRiD test set (RTX 4070 / Ryzen 5 7600X).}
\label{tab:15}
\scriptsize
\renewcommand{\arraystretch}{1.18}
\begin{tabularx}{\columnwidth}{@{}l*{1}{Y}@{}}
\toprule
\textbf{Operation} & \textbf{Time per image (ms)} \\
\midrule
ROI cropping & 0.06 \\
Green channel extraction & $\sim$0 \\
Median filtering & 14.08 \\
CLAHE & 32.90 \\
Tiling / resize & 932.52 \\
Tiled inference & 2067.17 \\
Confidence-threshold filtering & 0.90 \\
Ensembling & 24.21 \\
Inter-lesion suppression & 0.44 \\
\midrule
Total & 3072.29 \\
\bottomrule
\end{tabularx}
\end{table}

\FloatBarrier
\end{document}